\documentclass[aps,prd,twocolumn]{revtex4-1}

\usepackage{graphicx,amsfonts,amssymb,amsbsy}
\usepackage{pstricks}

\usepackage{graphicx}
\usepackage{slashed}
\usepackage{bm}
\usepackage{amssymb}

\begin{document}

\title{\bf Radial oscillations of color superconducting self-bound quark stars}
\author{C. V\'asquez Flores and G. Lugones}

\affiliation{Universidade Federal do ABC, Centro de Ci\^encias Naturais e Humanas, Rua Santa Ad\'elia, 166, 09210-170, Santo Andr\'e, Brazil.}

\vskip5mm
\begin{abstract}
We investigate the effect of the color-flavor locking pairing pattern on the adiabatic radial oscillations of pure self-bound quark stars using an equation of state in the framework of the MIT Bag model. We integrate the equations of relativistic radial oscillations to determine the fundamental and the first excited oscillation modes for several parameterizations of the equation of state.
For low mass stars we find that the period of the fundamental mode is typically $\sim 0.1$ ms and  has a small dependence on the parameters of the equation of state. For large mass stars the effect of color-flavor locking is related to the rise of the maximum mass with increasing $\Delta$. As for unpaired quark stars, the period of the fundamental mode becomes divergent at the maximum mass but now the divergence is shifted to large masses for large values of the pairing gap $\Delta$. As a consequence, the oscillation period is strongly affected by color superconductivity for stars with  $M \gtrsim 1.5 \; \textrm{M}_{\odot}$. We fit the period of the fundamental mode with appropriate analytical functions of the gravitational redshift of the star and the pairing gap $\Delta$.  We further discuss the excitation and damping of the modes and their potential detectability during violent transient phenomena.
\end{abstract}

\maketitle

\section{Introduction}

Neutron stars' oscillations  have been investigated for over almost 50 years (since the work of Chandrasekhar in 1964 \cite{Ref7}) using various equations of state (EOS) for nuclear matter and quark matter. The first exhaustive compilation of radial modes for various  zero  temperature  EOSs  was  presented  by  Glass \& Lindblom in 1983 \cite{Ref28}. Later studies by V\"ath \& Chanmugan \cite{Ref60} complemented and corrected earlier studies and also considered the case of strange quark stars. Further work by Kokkotas \& Ruoff \cite{Ref35} presented a new survey of the radial oscillation modes of neutron stars. Their study includes an extensive list of frequencies for the most common equations of state.  Radial pulsations were also considered in the case of newborn neutron stars (protoneutron stars) by Gondek et al. \cite{gondek}. They found that the spectrum of the lowest modes of radial pulsations of proto-neutron stars is quite different from that of cold neutron stars. Generally, protoneutron stars are significantly softer with respect to the radial pulsations, than cold neutron stars, and this difference increases for higher modes and for lower stellar masses \cite{gondek}. These differences stem from different structure of proto-neutron stars, which in contrast to cold neutron stars have extended envelopes, inflated by thermal and trapped neutrinos effects.

Radial oscillations were also studied in the case of quark stars. Stars containing quark phases fall into two main classes: hybrid stars (where quark matter is restricted to the core) and strange stars (made up completely by quark matter). It is expected that both kinds of stars cannot exist simultaneously in Nature, but it is not know which one would be realized (if any). This depends on whether the energy per baryon of $\beta$-equilibrated quark matter at zero pressure and zero temperature is less than the neutron mass (the so called ``absolute stability'' condition \cite{Ref1,fj84}). Analysis made within the MIT bag model shows that there is a room in the parameter space for the existence of strange stars. Moreover, color superconductivity enlarges substantially the region of the parameter space where $\beta$-stable quark matter has an energy per baryon smaller than the neutron mass \cite{Lugones2002,Lugones2003}. As a consequence, a ``color superconducting strange matter'' is allowed for the same parameters that would otherwise produce unbound strange matter \cite{Lugones2002}. On the other hand, within a Nambu-Jona-Lasinio (NJL) description of quark matter, the strange matter hypothesis is not favored, at least for the most accepted parameterizations of the EOS \cite{Buballa2005}. Thus, stars containing quark phases are believed to be hybrid stars within the NJL model.

Several works have dealt with the problem of radial oscillations of strange and hybrid stars \cite{Ref60,benvenuto1991,Ref14,Anand2000,Singh2002,gondek1999,Ref30}.
V\"ath \& Chanmugan \cite{Ref60} showed that the oscillation frequencies of strange stars have qualitatively a different dependence on the central density compared with the case of neutron stars, as the periods of all modes go to zero when the central density of the strange star approaches its smallest possible value. Benvenuto \& Horvath \cite{benvenuto1991} also presented calculations of radial oscillations of homogeneous strange stars, using a parameterized form of the equation of state. They showed that the particular form of the equation of state allows some simple and general scaling relations which may prove to be very useful for the search of these objects. Strange stars have been mostly studied in the framework of the MIT bag model, but some work has been made using other microphysical description of quark matter. In particular, Benvenuto and Lugones \cite{Ref14} studied the radial oscillations of strange stars in the Quark-Mass Density-Dependent model \cite{benvenuto1995a,benvenuto1995b} (QMDD model). Their results showed that oscillation periods are similar to those obtained within the MIT bag model. Recently, Anand et al. \cite{Anand2000} studied the radial oscillations of magnetized strange stars in the QMDD model showing that the squares of the frequencies are always decreasing functions of the central density of the strange star. In a similar way Singh et al. \cite{Singh2002} investigated the radial oscillations of rotating strange stars in strong magnetic fields in the QMDD model. They showed that the  difference in frequency between rotating and non-rotating stars is larger for higher magnetic fields. The change is small for low mass stars but it increases with the mass of the star. This change of frequency is significant for the most massive stars whereas it is marginal for a 1.4 M$_{\odot}$ star.

In the case of hybrid stars Gupta et al. \cite{Ref30} studied the effect of a mixed quark-nuclear matter core on radial oscillations. They found that the effect of the mixed phase is to decrease the maximum mass of the stable neutron star and to cause a kink in radial oscillation frequencies at the onset of the mixed phase. This kink can be traced to a slight kink in the density profile as well as in the EOS for neutron stars having a mixed quark-hadron phase  \cite{Ref30}.
Their results appears to be robust and independent of the EOS. Gondek and Zdunik \cite{gondek1999} also studied radial pulsations of neutron stars and strange quark stars with a nuclear crust. They used neutron star models constructed using a realistic equation of state of dense matter and strange star models using a phenomenological Bag model of quark matter. They calculated the eigenfrequencies of the three lowest modes of linear, adiabatic pulsations and found an avoided crossing phenomenon that is strongly related to the changes of compressibility of the matter throughout the star.

In this work we shall focus on the radial oscillations of strange quark stars paying particular attention to the effect of color superconductivity, which for the best of our knowledge, has not been taken into account yet in the literature. In particular, we shall consider the effect of color-flavor locking within the frame of the MIT bag model. The paper is organized as follows: in Sec. II we present the equation of state of the CFL phase and find simple analytical expressions for the relevant thermodynamical quantities. In Sec. III we study the radial oscillations of CFL strange stars and we calculate the fundamental and first excited modes for different values of the parameters of the EOS. In Sec. IV we discuss our results.

\section{Thermodynamics of the CFL phase} \label{EOS}

As mentioned before, we work here within the hypothesis that quark matter is absolutely stable and thus
quark stars are completely made up by an almost symmetric mixture of up, down and strange quarks. At sufficiently large densities and low temperatures quark matter is a color superconductor, which is a degenerate Fermi gas of quarks with a condensate of Cooper pairs near the Fermi surface \cite{BailinLove}. Color superconducting quark matter can come in a multiplicity of different phases, based on different pairing patterns of the quarks.  At asymptotically  large densities, where the quark masses are negligibly small compared to the quark chemical potential, three-flavor quark matter is in the color-flavor locked (CFL) state \cite{alford1999}. In this state quarks form Cooper pairs of different color and flavor where all quarks have the same Fermi momentum and electrons cannot be present \cite{Ref4}. Color-flavor locking has a profound effect on the properties of quark matter, mainly on transport properties such as mean free paths, conductivities and viscosities. Concerning the equations of state, the effects enter as a term of order $(\Delta/\mu)^2$ which is of a few percent for typical values of the color superconducting gap ($\Delta \sim 0-150$ MeV) and the baryon chemical potential ($\mu \sim 300-400$ MeV). However, the effect is proportionally very large in the low pressure regime that affects the absolute stability of quark matter. Thus, pure self-bound quark-matter stars (\textit{strange stars}) may exist for a wider range of parameters of the MIT Bag model equation of state \cite{Lugones2002}. This affects considerably the mass-radius relationship of CFL stars, allowing for very large maximum masses \cite{Lugones2003,Lugones2004}.

The equation of state for CFL quark matter can be obtained in the framework of the MIT bag model. To order $\Delta^{2}$, the thermodynamical potential $\Omega_{\rm CFL}$  reads \cite{Lugones2002}
\begin{equation}\label{OmegaCFL}
\Omega_{\rm CFL}= \Omega_{\rm free} - \frac{3}{\pi^{2}}\Delta^{2}\mu^{2} + B,
\end{equation}
being $\Omega_{\rm free}$ the thermodynamical potential of a state of unpaired
$u$, $d$ and $s$ quarks in which all them have a common Fermi momentum $\nu$, with $\nu$ chosen to minimize $\Omega_{\rm free}$. The binding energy of the diquark condensate is included by subtracting a condensation term proportional to $\Delta^{2}\mu^{2}$ where the chemical potential  $ \mu  \equiv  (\mu_u + \mu_d + \mu_s) / 3$ is related to $\nu$ through  $\nu = 2 \mu - ( \mu^2 + {m_s^2}/{3})^{1/2}$, being $m_s$ the mass of the strange quark. Confinement is introduced through a phenomenological vacuum energy density or bag constant $B$.

From the $\Omega_{\rm CFL}$ given above we can obtain the following expressions
for the pressure $p$ and the energy density $\epsilon$ to order $m^{2}_{s}$ \cite{Lugones2002}:
\begin{equation}\label{PB}
p=\frac{3\mu^{4}}{4\pi^{2}}+\frac{9\alpha \mu^{2}}{2\pi^{2}} - B,
\end{equation}

\begin{equation}\label{energiaB}
\epsilon = \frac{9\mu^{4}}{4\pi^{2}}+\frac{9\alpha\mu^{2}}{2\pi^{2}} + B,
\end{equation}
where
\begin{equation}\label{alfa}
\alpha = -\frac{m^{2}_{s}}{6}+\frac{2\Delta^{2}}{3}.
\end{equation}
In order to have the EOS in the form $\epsilon = \epsilon(p)$, we can invert Eq. (\ref{PB}) to find $\mu$ as a function of $p$
\begin{equation}\label{mu2}
\mu^{2}= -3\alpha +\bigg[\frac{4}{3}\pi^{2}(B+p)+9\alpha^{2}\bigg]^{1/2},
\end{equation}
and we can write $\epsilon=\epsilon(p)$ from Eqs. (\ref{PB}) and (\ref{energiaB})
\begin{equation}\label{energiaC}
\epsilon = 3p+4B-\frac{9\alpha\mu^{2}}{\pi^{2}},
\label{eq_ep}
\end{equation}
with $\alpha$ and $\mu$ given by Eqs. (\ref{alfa}) and (\ref{mu2}) respectively.

If we need $p=p(\epsilon)$ we may write
\begin{equation}\label{PC}
p=\frac{\epsilon}{3}-\frac{4B}{3}+\frac{3 \alpha\ \mu^{2}}{\pi^{2}},
\end{equation}
with $\mu$ given by
\begin{equation}\label{mu2B}
\mu^{2}=-\alpha + \bigg[\alpha^{2}+\frac{4}{9}\pi^{2}(\epsilon-B)\bigg]^{1/2}.
\end{equation}

For the oscillation equation (see Sec. III) we need the adiabatic index $\Gamma = (\epsilon+p)p^{-1}dp/d\epsilon$.
From Eq. (\ref{PC}) we have
\begin{equation}\label{dp_de}
\frac{dp}{d\epsilon}=\frac{1}{3}+\frac{6\alpha\mu}{\pi^{2}}\frac{d\mu}{d\epsilon}.
\end{equation}
The derivative $d\mu/d\varepsilon$ can be calculated from Eq. (\ref{mu2B}): ${d\mu}/{d\epsilon}={\pi^{2}}{(9\mu)^{-1} [\alpha^{2}+4\pi^{2}(\epsilon-B)/9]^{-1/2}}$. Thus, we have
\begin{eqnarray}\label{dmu_de}
\frac{dp}{d \epsilon} =\frac{1}{3}+\frac{2\alpha}{3}\bigg(\frac{1}{\mu^{2}+\alpha}\bigg),
\end{eqnarray}
which allows to write the adiabatic index $\Gamma$ as function of $p$ using Eq. (\ref{mu2})
or as a function of $\epsilon$ using Eq. (\ref{mu2B}).

Since the values of  $B$, $m_s$ and $\Delta$ are not accurately known we shall consider
them as free parameters in the equation of state. We emphasize that all the values
of  $B$, $m_s$ and  $\Delta$ employed in this paper
fall inside the \textit{stability windows} presented in Fig. 2 of Ref. \cite{Lugones2002}; i.e. we always
obtain \textit{strange stars} when integrating the stellar structure equations. Additionally, the parameters
satisfy the stability condition $m_s^2 < 2 \mu \Delta$ given in Ref. \cite{Alford2004}.

In our calculations of the next section we shall also employ the MIT equation of state for a gas of unpaired massless quarks, i.e. $\epsilon = 3 p + 4 B$.
Notice that unpaired matter is the same that CFL matter with $\Delta = 0$  only if all the quark masses are zero. If $m_s$ is not zero, the Fermi momenta of $u$, $d$ and $s$ quarks are different for unpaired quark matter, but they are still the same for CFL matter.

\section{Radial Pulsations of CFL stars}
In this section we shall study the radial oscillations of strange quark stars employing the equation of state of the previous section. We shall consider the unperturbed star to be composed of a perfect fluid, whose stress-energy tensor takes the form
\begin{equation}
{T}_{\mu\nu} =(\epsilon + p) u_{\mu}u_{\nu} + p{g}_{\mu\nu}.
\end{equation}

The generic background space-time of a static spherical star is expressed through the line element
\begin{equation}
\label{dsz_tov}
ds^{2}=-e^{ \nu(r)} dt^{2} + e^{ \lambda(r)} dr^{2} + r^{2}(d\theta^{2}+\sin^{2}{\theta}d\phi^{2}) .
\end{equation}
The Einstein equations in such a spacetime lead to the following set of stellar structure
equations (Tolman-Oppenheimer-Volkoff equations)
\begin{eqnarray}
\label{tov1}
&&\frac{dp}{dr} = - \frac{\epsilon m}{r^2}\bigg(1 + \frac{p}{\epsilon}\bigg)
	\bigg(1 + \frac{4\pi p r^3}{m}\bigg)\bigg(1 -
	\frac{2m}{r}\bigg)^{-1},
\\ \nonumber \\
\label{tov2}
&&\frac{d\nu}{dr} = - \frac{2}{\epsilon} \frac{dp}{dr}
	\bigg(1 + \frac{p}{\epsilon}\bigg)^{-1},
\\ \nonumber \\
\label{tov3}
&&\frac{dm}{dr} = 4 \pi r^2 \epsilon,
\end{eqnarray}
where $m$ is the gravitational mass inside the radius $r$. The metric function $\nu$ has the boundary condition
\begin{equation}
\label{BoundaryConditionMetricFunction}
    \nu(r=R)= \ln \bigg( 1-\frac{2M}{R} \bigg),
\end{equation}
where $R$ is the radius of the star and $M$ its mass. With this condition the  metric function $\nu(r)$ will match smoothly to the Schwarzschild metric outside the star.

To obtain the equations that govern radial oscillations, both
fluid and spacetime variables are perturbed in such a way that the
spherical symmetry of the background body is not violated. These perturbations
are inserted into the Einstein equations and into the energy, momentum and
baryon number conservation equations and only the first-order terms are retained.

Several forms of the oscillation equation have been presented in the literature.
The original form presented by Chandrasekhar \cite{Ref7} constitutes a Sturm-Lioville problem
whose solution provides the eigenvalues and the eigenfunctions for the radial perturbations
(see eq. (59) of Ref. \cite{Ref7}). The second order oscillation equation given by Chandrasekhar can
be split into two first order equations. This has been done by Vath and Chanmugan \cite{Ref60} who
derived a set of first order equations for the quantities $\Delta r/r $ and $\Delta p/p$ (see eqs. (9) and (10) of Ref. \cite{Ref60}). More recently, Gondek et al. \cite{gondek} obtained a similar set
of equations but for the relative radial displacement $\Delta r/r $ and the Lagrangian perturbation of
the pressure $\Delta p$ (see eqs. (11) and (12) of Ref. \cite{gondek}). All these sets of equations
are equivalent. In this work, we adopted the equations of Gondek et al. \cite{gondek}, because this
system is particularly suitable for numerical applications and the boundary condition at the star's
surface can be obtained by purely physical arguments. Another important advantage of this system of
oscillation equations stems from the fact that they do not involve any derivatives of the adiabatic
index, $\Gamma$. Adopting $G = c = 1$ the system of equations is
\begin{equation}\label{ecuacionparaXI}
\frac{d\xi}{dr}=-\frac{1}{r}\bigg(3\xi+\frac{\Delta p}{\Gamma p}\bigg)-\frac{dp}{dr}\frac{\xi}{(p+\epsilon)},
\end{equation}
\begin{eqnarray}\label{ecuacionparaP}
\frac{d\Delta p}{dr}=\xi \bigg\{\omega^{2}e^{\lambda-\nu}(p+\epsilon)r-4\frac{dp}{dr} \bigg \} \nonumber
\\ \nonumber \\
+\xi\bigg \{\bigg(\frac{dp}{dr}\bigg)^{2}\frac{r}{(p+\epsilon)}-8\pi e^{\lambda}(p+\epsilon)pr \bigg \}
\\ \nonumber \\
 +\Delta p\bigg \{\frac{dp}{dr}\frac{1}{(p+\epsilon)}-
4\pi(p+\epsilon)r e^{\lambda}\bigg\},\nonumber
\end{eqnarray}
where $\omega$ is the eigenfrequency and the quantities $\xi \equiv \Delta r / r$
and $\Delta p$ are assumed to have a harmonic time dependence $\varpropto e^{i\omega t}$.

%
\begin{figure*}[htp]
 \centering
\includegraphics[angle=0,scale=0.27]{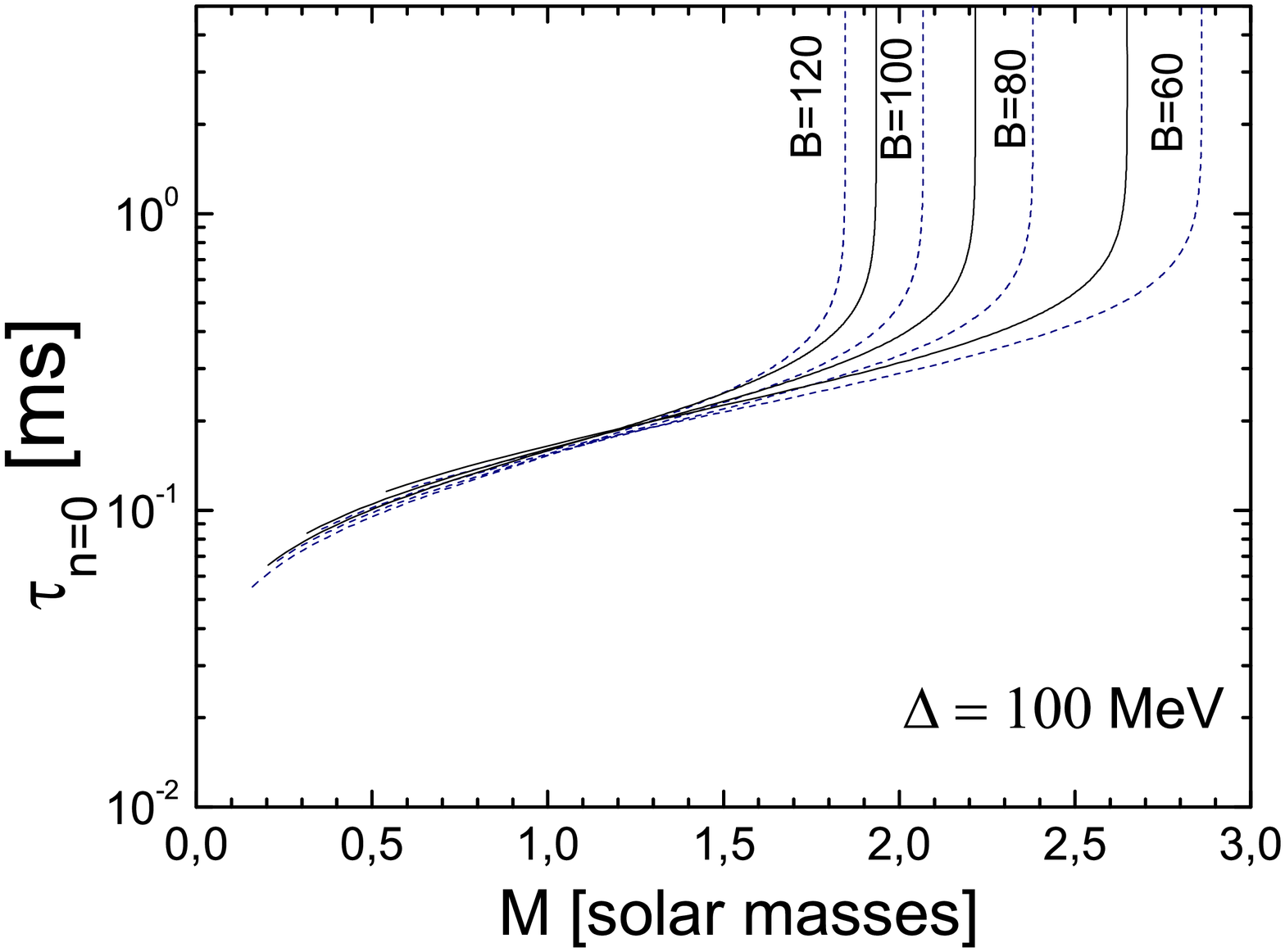}
\includegraphics[angle=0,scale=0.27]{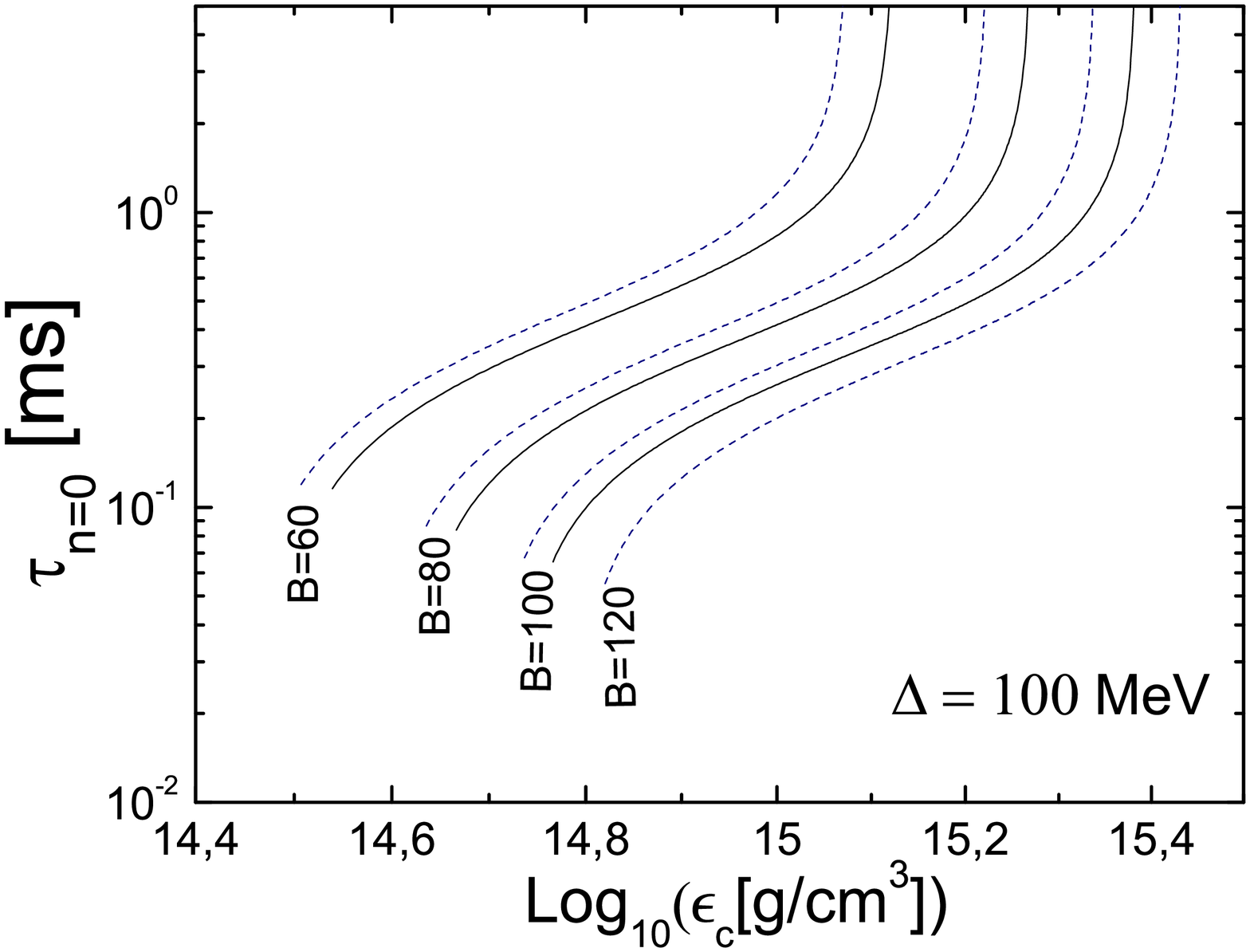}
\includegraphics[angle=0,scale=0.27]{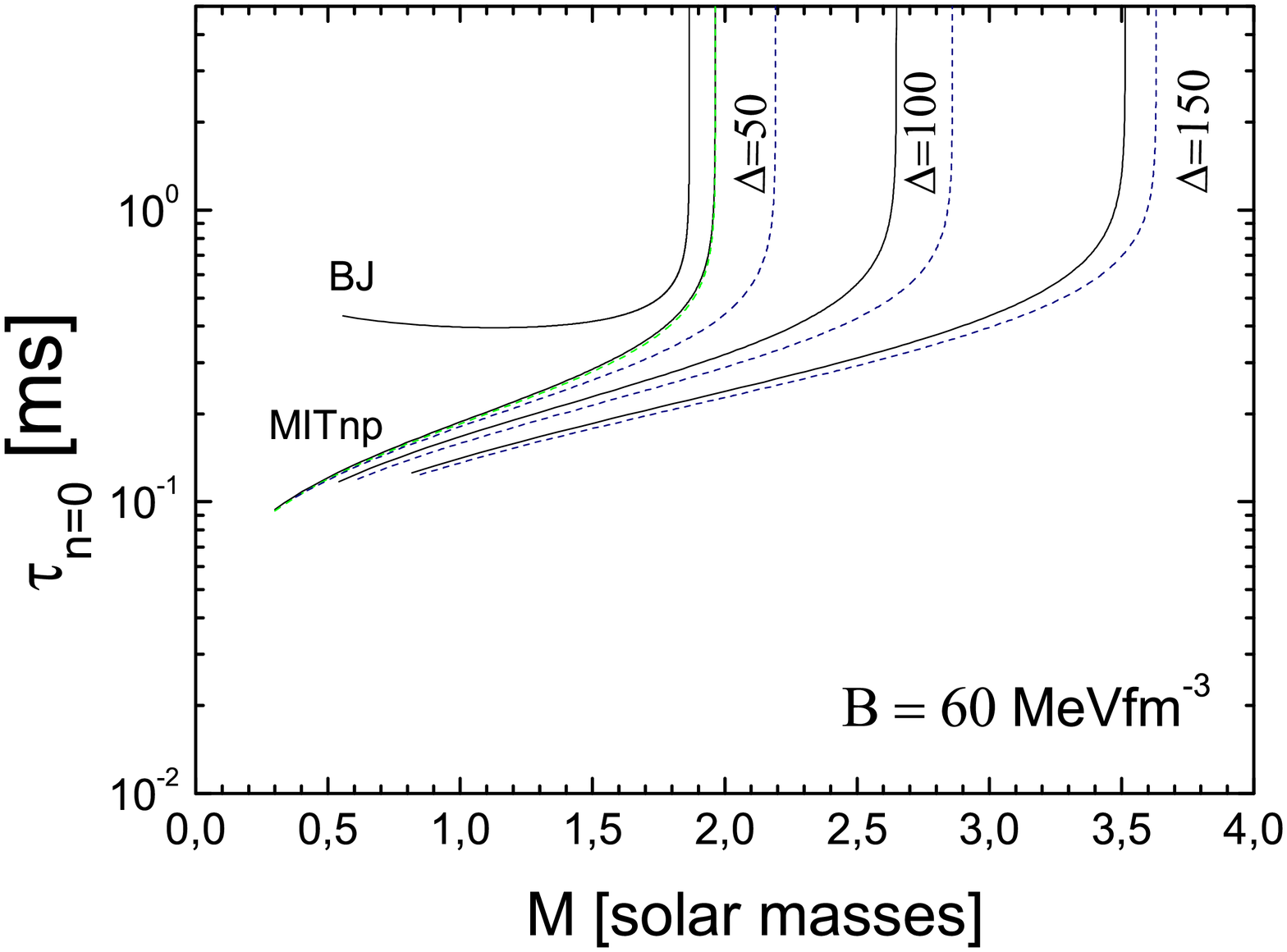}
\includegraphics[angle=0,scale=0.27]{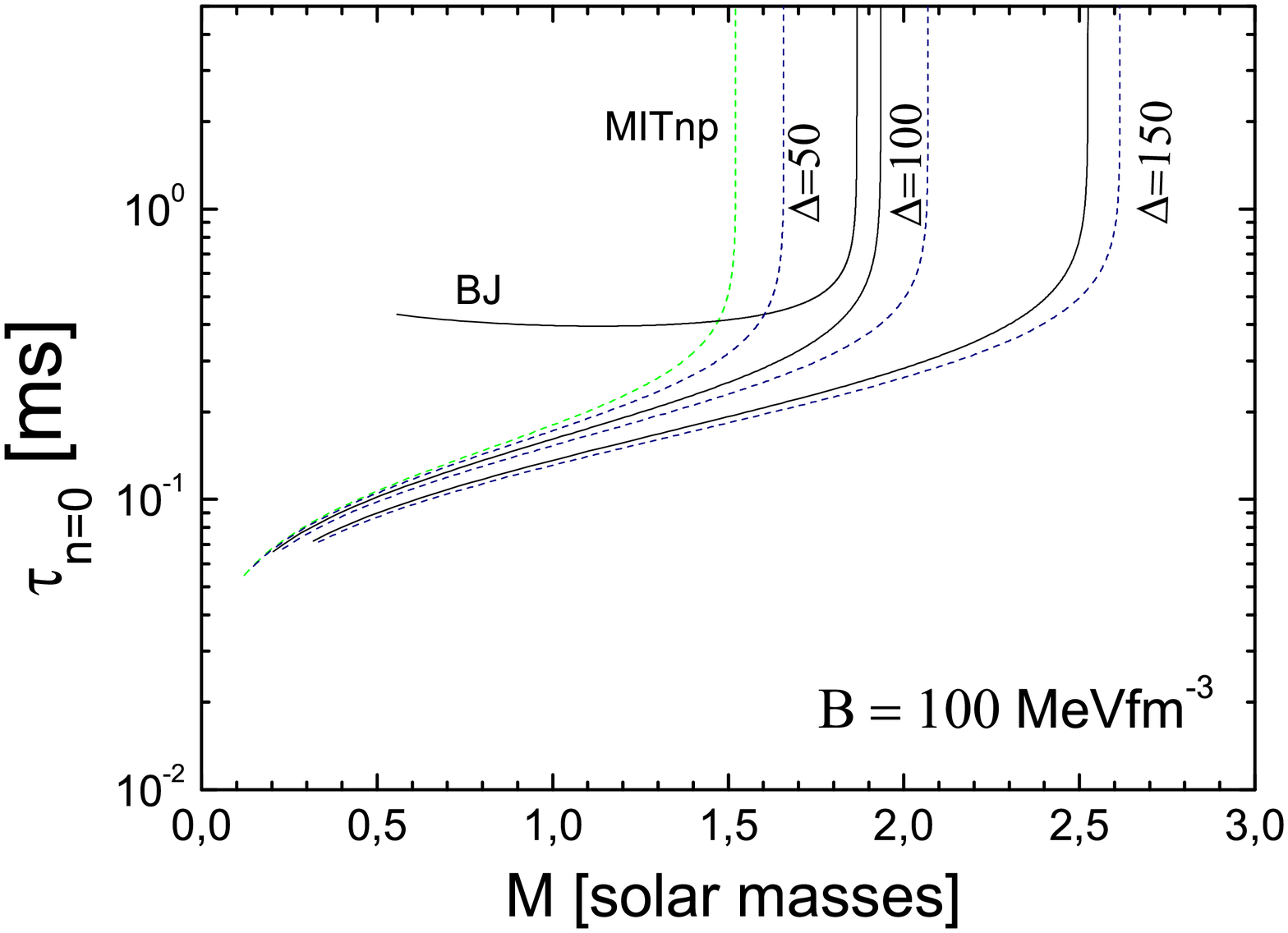}
\caption[fig1]{\begin{footnotesize}  The period of the fundamental oscillation mode (in milliseconds)
as a function of the mass $M$ and the central energy density $\epsilon_c$ of the star.
In the upper panels the results are shown for CFL strange stars with $\Delta=100$ MeV
and different values of the bag constant $B$ and the strange quark  mass  $m_{s}$.
$B$ is given in MeV fm$^{-3}$ and $m_{s}$ in MeV.  Dashed lines correspond
to $m_{s} = 0$ and solid lines to $m_{s}=$150 MeV. In the lower panels the results
are shown for hadronic stars employing the Bethe-Johnson EOS (BJ), for strange stars
without color superconductivity (MITnp), and  for CFL strange stars with
$\Delta = 50, 100, 150$ MeV. The curve MITnp is given only for $m_{s}= 0$. In the left lower panel we fix the value of $B$ to 60 MeV fm$^{-3}$ and in the right
lower panel to 100 MeV fm$^{-3}$.
Comparing the curve MITnp with the CFL curves for $\Delta = 50, 100, 150$ MeV we notice that the effect of color superconductivity is large for stars
with $M \gtrsim 1.5 \; \textrm{M}_{\odot}$. In particular, notice that the curve with $\Delta = 50$ MeV and  $m_s = 150$ MeV is almost coincident with
the curve MITnp with $m_{s}= 0$ (see left lower panel). This happens because the constant $\alpha$ in Eq. (4) is very small in the former case and it
is zero in the latter. Thus, the equation of state given in Eq. (6) is very similar in both cases.  If we compare the curve for $\Delta = 50$ MeV and for MITnp both having the same $m_s$ we see that the effect of color superconductivity is non negligible for $\Delta = 50$ MeV.
The curve for $\Delta = 50$ MeV and $m_s = 150$ MeV is not shown in the right lower panel because the parameters of the equation of state fall outside the stability window given in Ref. \cite{Lugones2002}.
\end{footnotesize}}
\label{fig1}
\end{figure*}

\begin{figure*}[htp]
 \centering
\includegraphics[angle=0,scale=0.27]{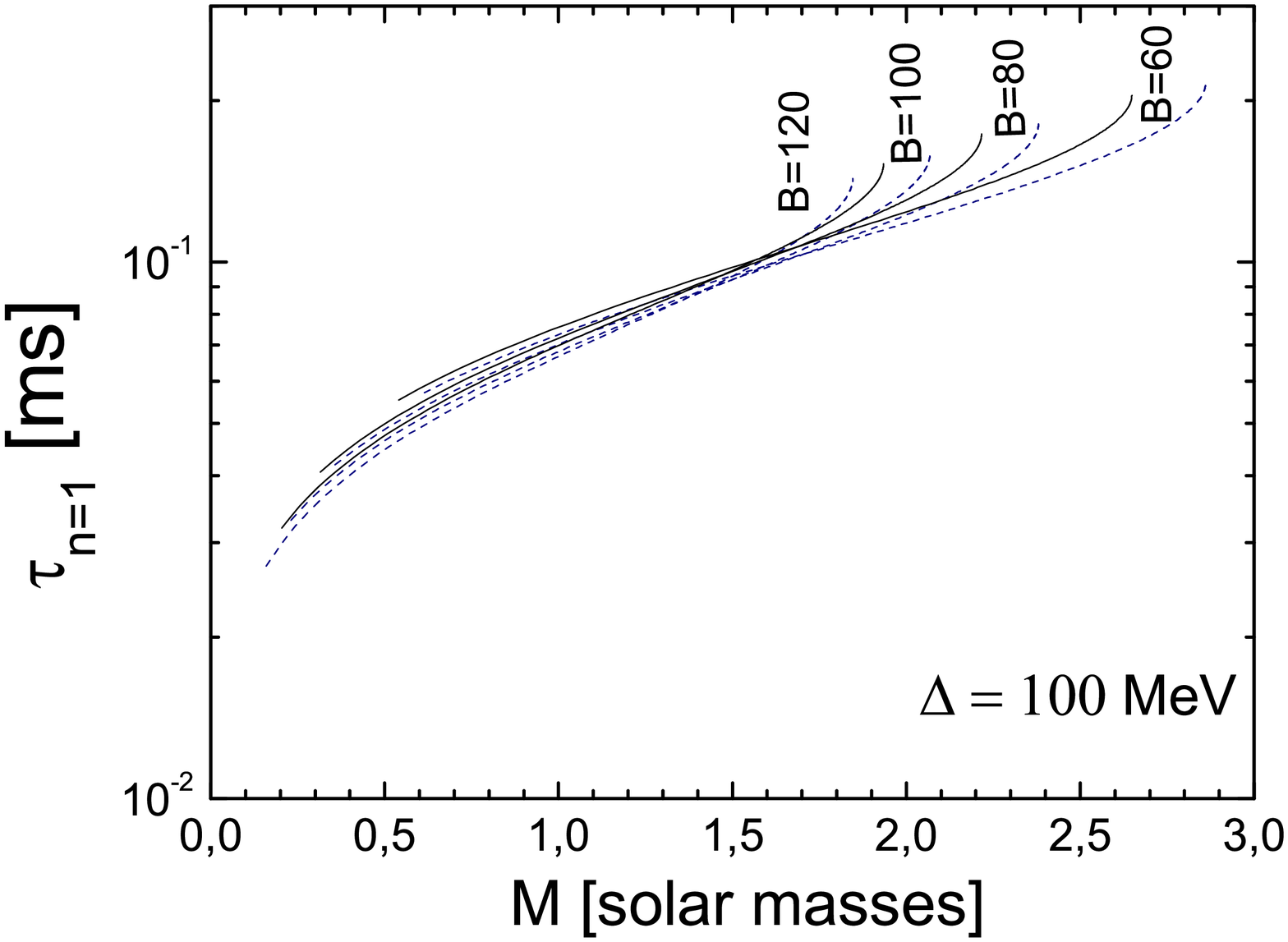}
\includegraphics[angle=0,scale=0.27]{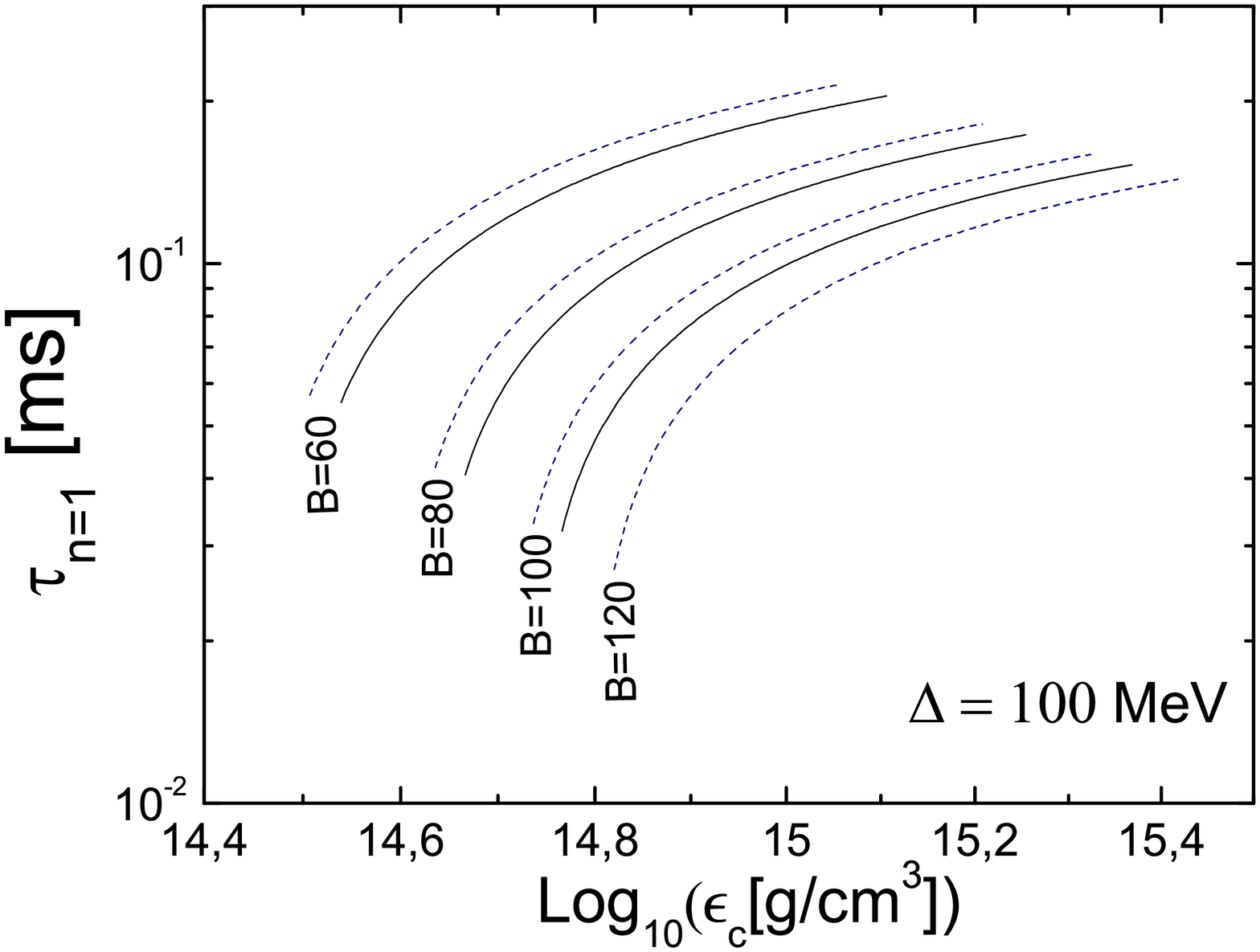}
\includegraphics[angle=0,scale=0.27]{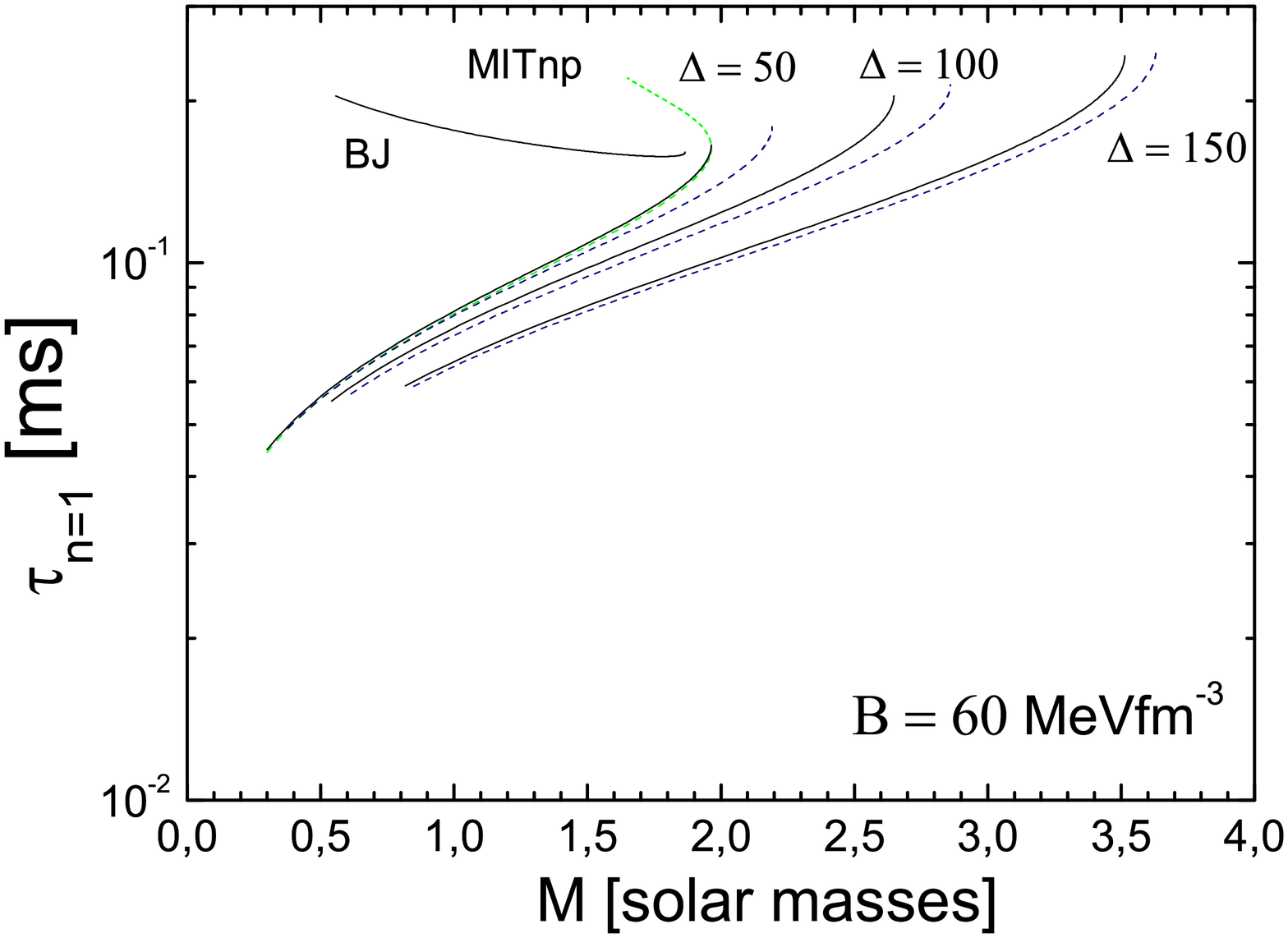}
\includegraphics[angle=0,scale=0.27]{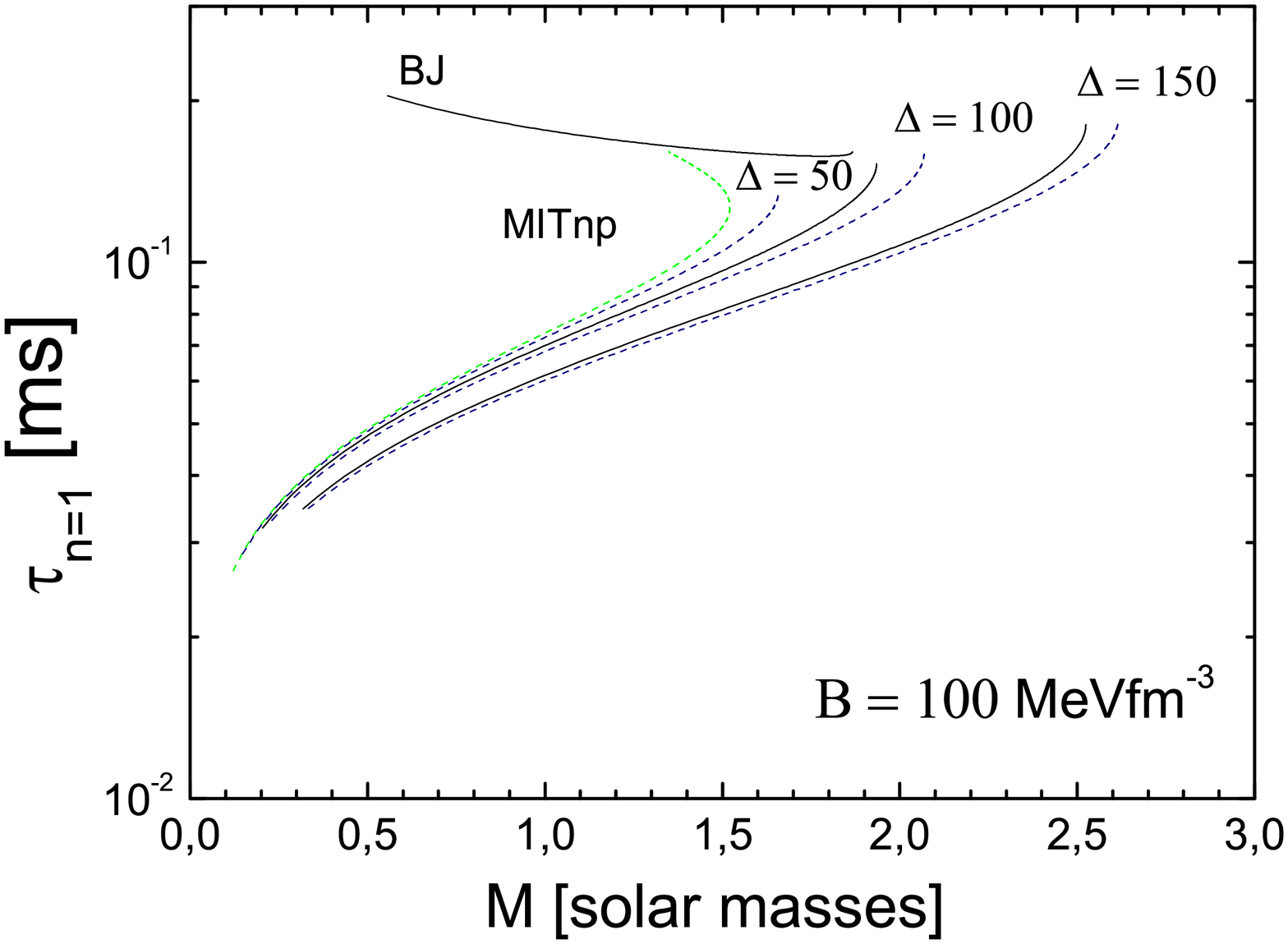}
\caption[fig2]{\begin{footnotesize}
Same as Fig. 1 but for the period of the first excited mode.
\end{footnotesize}}
\label{fig2}
\end{figure*}

To solve equations (\ref{ecuacionparaXI}) and (\ref{ecuacionparaP}) one needs two boundary conditions.
The  condition of regularity at $r=0$ requires that for $r \rightarrow 0$ the coefficient of the $1/r$
term in Eq. (\ref{ecuacionparaXI}) must vanish \cite{Ref60,gondek,gondek1999}. Thus, we have
\begin{equation}\label{DeltaP}
(\Delta p)_{center}=-3(\xi \Gamma p)_{center}.
\end{equation}
Notice that the eigenfunctions can be normalized in order to have $\xi(0)=1$. The surface of the star is determined by the condition that for $r\longrightarrow R$, one has $p \longrightarrow 0$. This implies that the Lagrangian perturbation in the pressure at the surface is zero. Therefore the second boundary condition is
\begin{equation}\label{PenSuperficie}
(\Delta p)_{surface}=0.
\end{equation}

In order to numerically solve the oscillation equations we proceed as follows.
First, we integrate the Tolman-Oppenheimer-Volkoff equations for each set of the parameters
of the equation of state ($B$,  $m_s$ and  $\Delta$)  in order to obtain the coefficients
of the oscillation equations for a given central pressure.
Then we solve the oscillation equations by means of the  {\em shooting method}:
we start the numerical integration of Eqs. (\ref{ecuacionparaXI}) and (\ref{ecuacionparaP}) for a
trial value of $\omega^2$ and a given set of initial values of
$\xi(r=0)$ and $\Delta p(r=0)$ which satisfy at the center the boundary
condition given above. The equations are integrated outwards trying to
match the boundary condition at the star's surface. After each integration, the
trial value of $\omega^2$ is corrected in order to improve the matching of the
surface boundary condition until the desired precision is achieved.
The discrete values of $\omega^2$ for which Eq.
(\ref{PenSuperficie}) is satisfied are the eigenfrequencies of the radial
perturbations. Our code was able to reproduce the results of Vath \& Chanmugam \cite{Ref60}
and of Kokkotas \& Ruoff \cite{Ref35}. 	

Our results are shown in Figs. 1$-$4 for different values of the parameters $B$, $m_s$ and $\Delta$ of the equation of state (falling inside the \textit{stability windows} presented in Fig. 2 of Ref. \cite{Lugones2002}). In Figs. 1 and 2 we show the period of the fundamental and the first excited modes of CFL strange stars as a function of the stellar mass $M$  and the central energy density $\epsilon_c$. The effect of color superconductivity has two aspects. By one hand, for a fixed mass of the star  the period of the fundamental mode is smaller as $\Delta$ increases. On the other hand, the maximum mass of the star increases significantly for large values of  $\Delta$. As a consequence, the oscillation period is largely affected by color superconductivity for stars with masses near the maximum mass. In Figs. 3 and 4 we show the period of the fundamental and the first excited modes as a function of the gravitational redshift $z$ at the surface of the compact star. The effect of $\Delta$ is similar to the observed in the plots of $\tau$ versus $M$ (c.f. Figs. \ref{fig1} and \ref{fig2}).
For comparison, the calculations were also performed for hadronic stars described by the Bethe-Johnson EOS \cite{Shapiro} and for quark stars without color superconductivity. For low mass stars there is a large difference between the oscillation periods of hadronic stars and quark stars, as already known from previous calculations without color superconductivity (see lower panels of Figs. \ref{fig1} and \ref{fig2}). The difference is also large for high mass stars but this is due to the difference between the maximum mass in different models. A similar behavior is found in the plots of $\tau$ versus $z$ (see Figs. 3 and 4).

Notice that while other EOS parameters such as  $B$ or $m_s$ can have similar effects than $\Delta$ on the oscillation periods, the effect of color superconductivity is stronger for $\Delta \sim 100$ MeV or larger. In fact, $m_s$ and $\Delta$ only enter in the equation of state through the parameter $\alpha$ defined in Eq. (\ref{alfa}) and they have opposite effects. However, since $m_s^2$ is multiplied by $1/6$ and $\Delta^2$ by $2/3$, the effect of $m_s$ is proportionally smaller that the effect of  $\Delta$. In the case of $B$, we notice that the period of the fundamental mode of stars with  $M \lesssim 1.5 \; \textrm{M}_{\odot}$ is rather independent of the value of $B$ (and also $m_s$) but changes up to a factor of $\sim 2$ in the here-considered range of $\Delta$.
It is also interesting to note that, for equations of state of the form $P =  (\epsilon - 4 B)/3$,  it has been shown that there exists a simple scaling law relating the oscillation periods corresponding to different values of $B$. The scaling law is $\tau_n^{\prime} = (B/B^{\prime} )^{1/2} \times \tau_n$ where the oscillation period $\tau_n^{\prime}$ corresponds to $B^\prime$ and $\tau_n$ corresponds to $B$ (see \cite{benvenuto1991} for more details). Thus, for that equation of state, there is shifting of the periods for different values of $B$. In our case, a similar shifting is observed in the left upper panels of Figs. 1 and 2 but it is not governed by a simple scaling law like the one given above because of the presence the condensation term in Eq. (\ref{PC}), which is not a constant but depends on the energy density $\epsilon$.

\begin{figure*}[htp]
 \centering
\includegraphics[angle=0,scale=0.27]{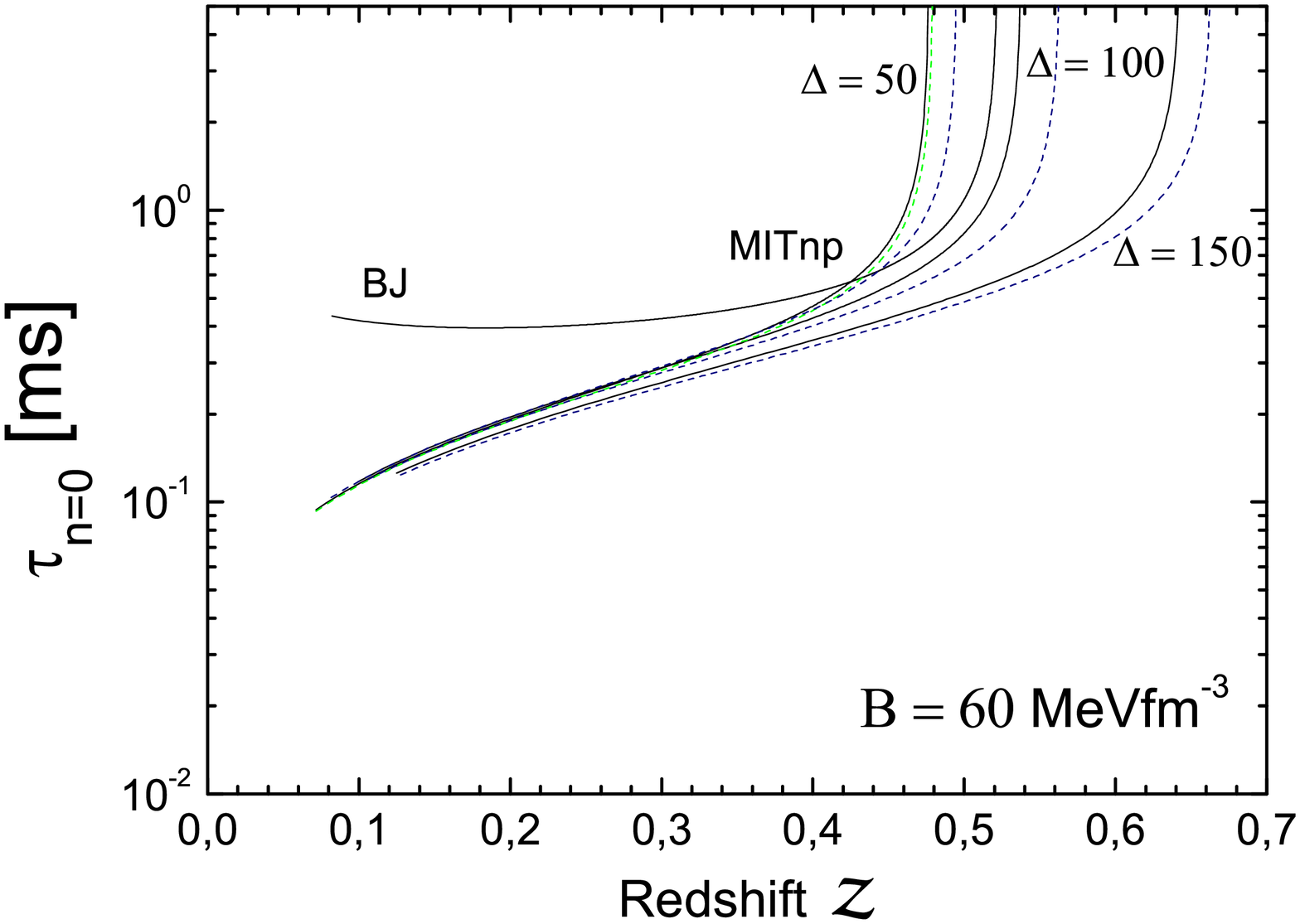}
\includegraphics[angle=0,scale=0.27]{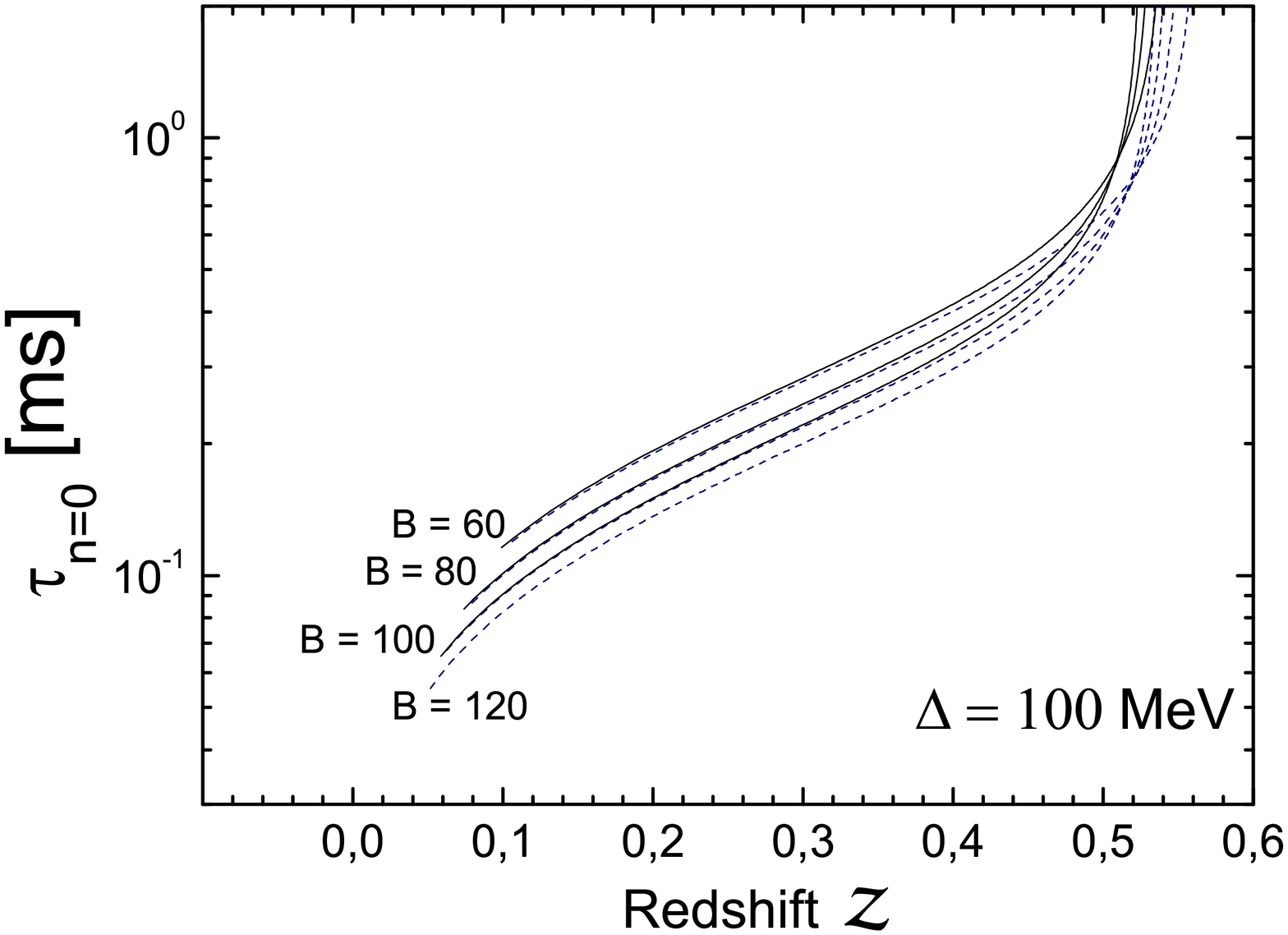}
\caption[fig2]{\begin{footnotesize}
The period for the fundamental oscillation mode as a function of the gravitational redshift $z$ at the surface of the CFL strange star. In the left panel the results are shown for CFL strange stars with $B$ = 60 MeV fm$^{-3}$, $m_{s}$ = 150 MeV
and different values of  $\Delta$. We also show the curves for hadronic stars employing the
Bethe-Johnson EOS (BJ) and for strange stars without color superconductivity (MITnp).
In the right panel the results are shown for CFL strange stars with $\Delta =$ 100 MeV and different
values of $B$ (dashed lines correspond to $m_{s} = 0$ and solid lines to $m_{s} = $ 150 MeV).
\end{footnotesize}}
\label{fig3}
\end{figure*}

\begin{figure*}[htp]
 \centering
\includegraphics[angle=0,scale=0.27]{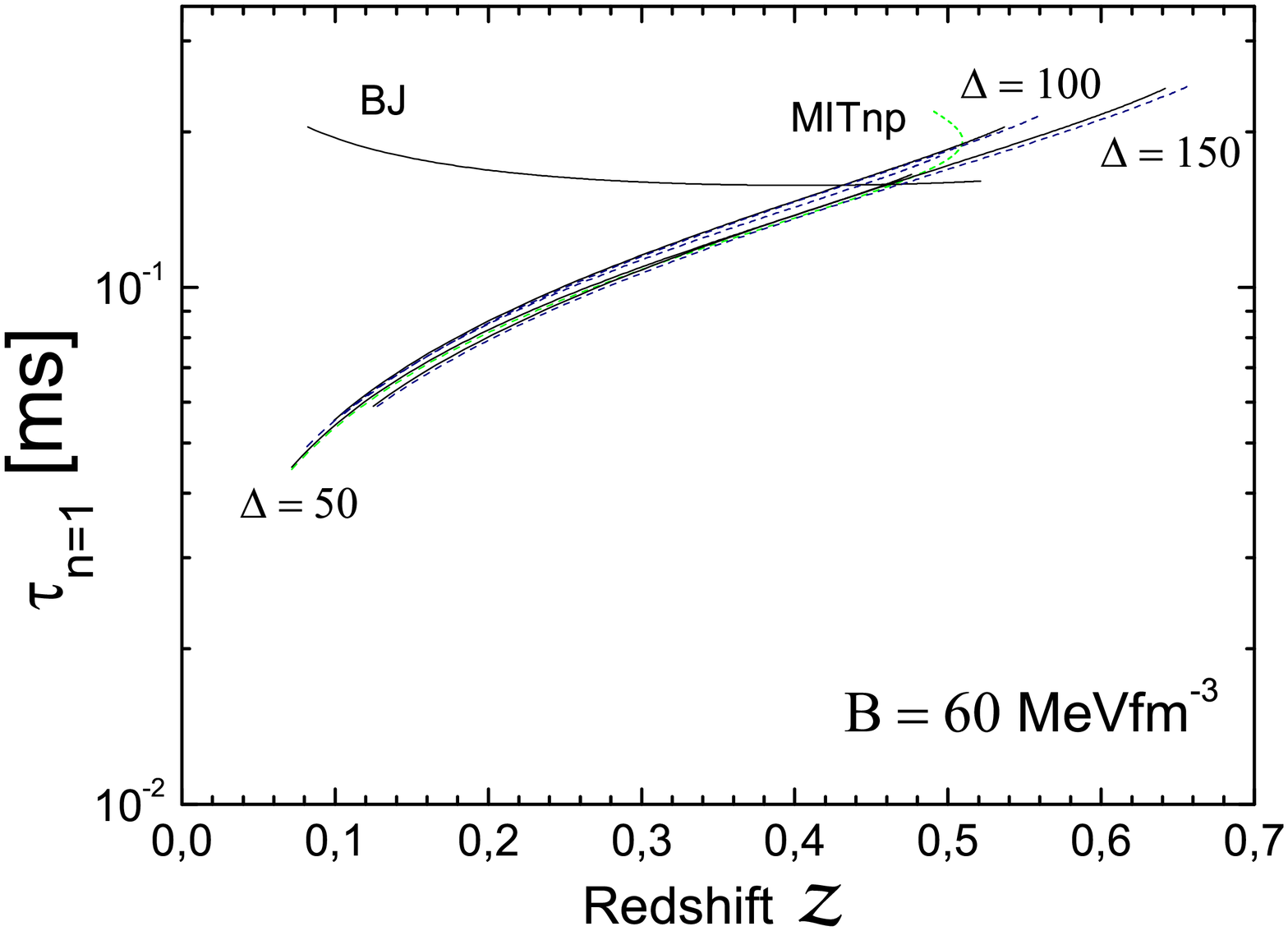}
\includegraphics[angle=0,scale=0.27]{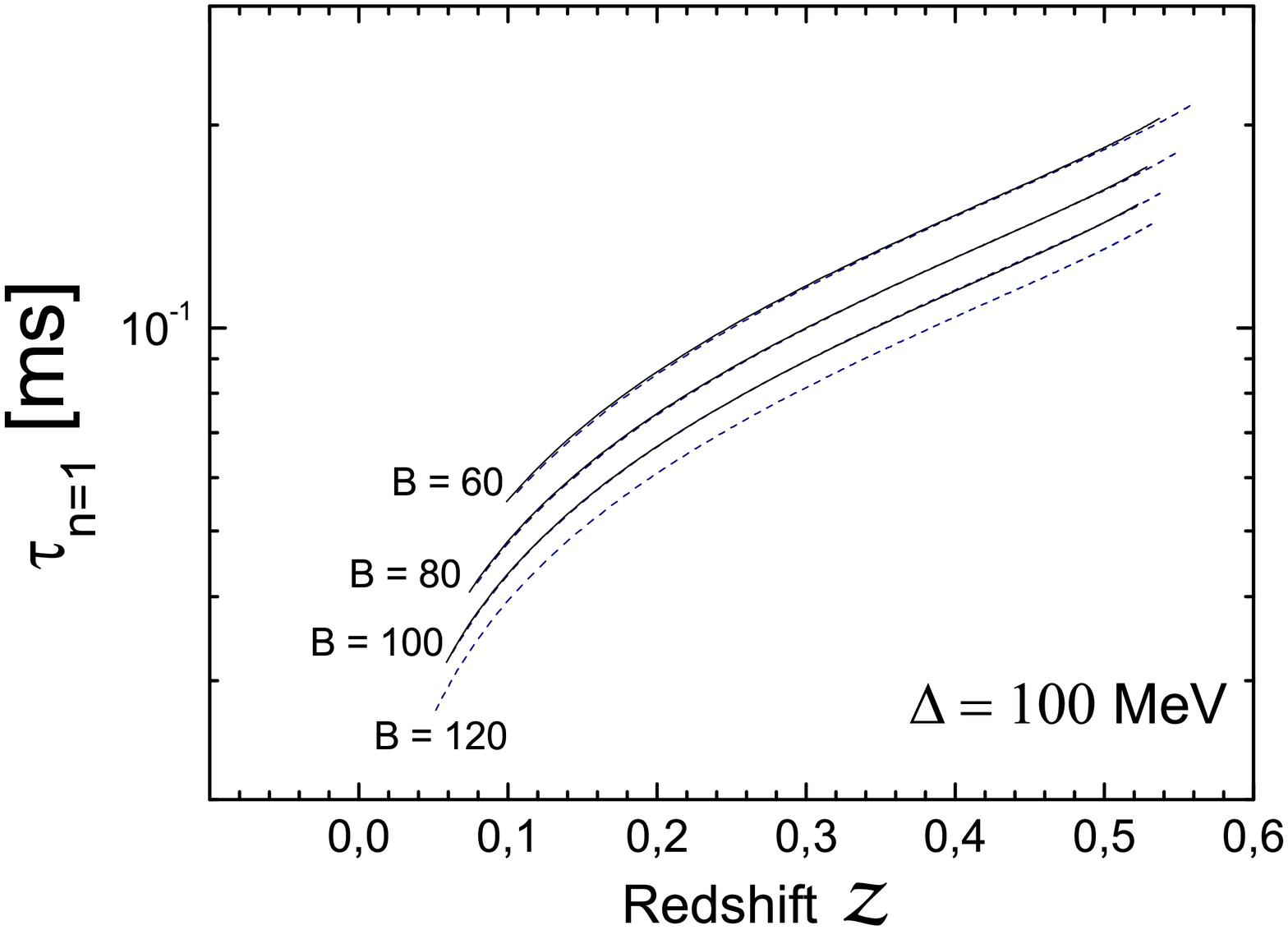}
\caption[fig2]{\begin{footnotesize}
Same as Fig. 3 but for the period of the first excited oscillation mode.
\end{footnotesize}}
\label{fig4}
\end{figure*}

For applications it is convenient to have analytic fittings of the results. The period of the fundamental mode can be expressed
as a function of the redshift $z$ of the star and the pairing gap $\Delta$ as:
\begin{equation}\label{taufit}
\tau_{n=0} = \frac{a_1 + a_{2} \delta + a_{3} z + a_{4} z^2 + a_{5} z \delta}{1 + b_1 \delta + b_2 z + b_3 \delta^2 + b_4 z^2 + b_5 z \delta},
\end{equation}
where $\tau_{n=0}$ is in milliseconds and $\delta \equiv \Delta$ / (100 MeV).
The coefficients are given in Table 1 for different values of the bag constant $B$.
This fitting is better than a $5 \%$ in the range  0 MeV $<$ $\Delta$ $<$  150 MeV and  $0.1  < z < 0.98 \times z_{max}$. The maximum gravitational redshift of the star $z_{max}$ as a function of $B$ and $\Delta$ can be obtained through
\begin{equation}\label{mmax}
z_{max} = c_0 + c_1 \Delta + c_2  B + c_3 \Delta^2 + c_4 B^2 + c_5 B \, \Delta ,		
\end{equation}
with  $B$ in MeV fm$^{-3}$  and $\Delta$ in MeV. The coefficients are given in  Table 2.
This fitting is better than a $5 \%$ for  0 MeV $<$ $\Delta$ $<$ 150 MeV and  $60 \, \textrm{MeV fm}^{-3} < B < 120 \, \textrm{MeV fm}^{-3}$.

\begin{table*}[t!]
\centering
\begin{tabular}{c||c|c|c|c}
\hline  \hline
&   \quad B =  60  MeV fm$^{-3}$  \quad  &  \quad   B = 80 MeV fm$^{-3}$  \quad  & \quad       B =  100  MeV fm$^{-3}$  \quad & \quad      B =    120 MeV fm$^{-3}$ \quad \\
\hline  \hline
$a_1$ &  6.04684 $\times 10^{-2}$  &  5.21480 $\times 10^{-2}$ &  4.59254 $\times 10^{-2}$  &  4.12120 $\times 10^{-2}$  \\ \hline
$a_2$ & -4.39575 $\times 10^{-2}$  & -2.20430 $\times 10^{-2}$ & -1.86426 $\times 10^{-2}$  & -1.55666 $\times 10^{-2}$  \\ \hline
$a_3$ &  5.06630 $\times 10^{-1}$  &  3.81115 $\times 10^{-1}$ &  3.39918 $\times 10^{-1}$  &  3.09221 $\times 10^{-1}$  \\ \hline
$a_4$ & -9.53275 $\times 10^{-1}$  & -7.82398 $\times 10^{-1}$ & -6.88127 $\times 10^{-1}$  & -6.26093 $\times 10^{-1}$  \\ \hline
$a_5$ &  2.10216 $\times 10^{-1}$  &  1.20291 $\times 10^{-1}$ &  8.61653 $\times 10^{-2}$  &  6.49367 $\times 10^{-2}$  \\ \hline
$b_1$ & -2.85725 $\times 10^{-1}$  & -1.73731 $\times 10^{-1}$ & -1.82305 $\times 10^{-1}$  & -1.77994 $\times 10^{-1}$  \\ \hline
$b_2$ & -1.25543                   & -1.81158                  & -1.85871                   & -1.92614                   \\ \hline
$b_3$ &  7.81002 $\times 10^{-2}$  &  5.28224 $\times 10^{-2}$ &  4.17669 $\times 10^{-2}$  &  3.44168 $\times 10^{-2}$  \\ \hline
$b_4$ & -1.59488                   & -4.63609 $\times 10^{-1}$ & -3.34299 $\times 10^{-1}$  & -1.82102 $\times 10^{-1}$  \\ \hline
$b_5$ &  8.21011 $\times 10^{-1}$  &  5.2234  $\times 10^{-1}$ &  5.08038 $\times 10^{-1}$  &  4.77180 $\times 10^{-1}$  \\ \hline
\end{tabular}
\caption{Coefficients of the polynomial fitting of the fundamental period as a function of the gravitational redshift of the star $z_{max}$ and the parameter $\delta$ (being  $\delta = \Delta$ / (100 MeV)), see Eq. (\ref{taufit}). This fitting was calculated for $B$ = 60,  80,  100,  120 MeV fm$^{-3}$. In all cases we assumed  $m_s = 150$ MeV.} \label{sets}
\end{table*}

\begin{table*}[t!]
\centering
\begin{tabular}{c|c|c|c|c|c}
\hline \hline
$c_0$ & $c_1$ &  $c_2$  & $c_3$ & $c_4$ & $c_5$ \\
\hline \hline
 4.60362$\times 10^{-1}$  \quad  & \quad  3.62306 $\times 10^{-4}$  \quad  & \quad  -8.89339 $\times 10^{-5}$
  \quad  & \quad 7.93274  $\times 10^{-6}$  \quad  & \quad  1.74139 $\times 10^{-6}$  \quad  & \quad -5.88166
 $\times 10^{-6}$ \quad \\
\hline
\end{tabular}
\caption{Coefficients of the polynomial fitting of the gravitational redshift of the star $z_{max}$ as a function of the parameters $B$ and $\Delta$.
As shown in Eq. (\ref{mmax}) we have $z_{max} = c_0 + c_1 \Delta + c_2  B + c_3 \Delta^2 + c_4 B^2 + c_5 B \Delta$, with $B$ in MeV fm$^{-3}$  and $\Delta$ in MeV. We assumed  $m_s = 150$ MeV.} \label{sets}
\end{table*}


\section{Discussion}

In this work we have presented a study of the radial pulsational properties of strange quark stars, paying particular attention to the effect of color superconductivity. We have shown  that the effect of color flavor locking into the oscillation periods is different for low mass stars (with $M \lesssim 1.5 \; \textrm{M}_{\odot}$) and for large mass stars (with $M \gtrsim 1.5 \; \textrm{M}_{\odot}$).

For low mass stars the period of the fundamental mode $\tau_{n=0}$ is typically $\sim 0.1$ ms. For a fixed $M$,  $\tau_{n=0}$ is almost independent on the value of $B$ and $m_s$ but decreases up to a factor of $\sim 2$ as $\Delta$ goes from 0 to $\sim 100$ MeV.

For large mass stars the effect of color flavor locking is related to the rise of the maximum mass with increasing $\Delta$. As for unpaired quark stars, $\tau_{n=0}$ becomes divergent at the maximum mass but now the divergence is shifted to large masses for large values of  $\Delta$. As a consequence, the oscillation period is strongly affected by color superconductivity for $M \gtrsim 1.5 \; \textrm{M}_{\odot}$.

It is well know that the periods of radial oscillations of strange stars behave very differently from those of hadronic stars \cite{Ref60}, specially for low mass stars.  This difference is amplified by color superconductivity because the oscillation periods tend to be smaller as the pairing gap $\Delta$ increases. Also, as for unpaired strange stars, the oscillation periods go to zero when the central energy density moves towards the smallest possible value $\epsilon_{min}$  for which the pressure is zero (see Eq. (\ref{eq_ep})). This can be understood by noticing that low-mass quark stars are very well described as non-relativistic constant-density spheres, which have oscillation periods proportional to the adiabatic index $\Gamma$ (for the fundamental mode we have $\omega_0^2 = 4 \pi G \rho (4 \Gamma - 3) / 3$ \cite{Shapiro}). Since the pressure tends to zero everywhere for a quark star with $\epsilon \sim \epsilon_{min} $, the adiabatic index $\Gamma = (\epsilon+p)p^{-1}dp/d\epsilon$ tends to infinity, $\omega^2$ diverges, and the oscillation period tends to zero. This behavior is clear in the right upper panels of Figs. 1 and 2.

In addition to knowing the spectrum of pulsations for a given internal composition, it is important
to realize whether the resulting modes are able to survive for a sufficiently long time provided they are excited in any astrophysically realistic situation. In principle, radial pulsations could be excited during a variety of catastrophic events taking place during the compact star's life. Starquakes and stellar collisions are potential mechanisms if
enough energy is transferred to vibrational modes.
However, it has been shown that radial pulsations of quark stars are quickly damped  due to the enormous bulk viscosity of  hot unpaired quark matter \cite{WangLu1984,Sawyer1989}.  The volume oscillation forces the system out of chemical equilibrium with respect to the non-leptonic process $u +d \leftrightarrow u + s$ and the semi-leptonic processes $u + e^- \leftrightarrow d + \nu_e$ and $u + e^- \leftrightarrow s + \nu_e$. In general, semi-leptonic processes are slower than the non-leptonic one, and the most effective damping reaction in unpaired quark matter is $u +d \leftrightarrow u + s$ \cite{WangLu1984,madsen1993}. Since $u$ and $d$ quarks are essentially massless while $m_s \sim 150$ MeV, the forward rate of $u +d \leftrightarrow u + s$ cannot keep equal to the reverse rate and the system cannot keep in equilibrium during oscillations. This leads to irreversible processes and damping in unpaired quark matter. For a typical stellar oscillation time of 10$^{-3}$ s, high amplitude oscillations are damped in fractions of a second due to the non-leptonic process \cite{madsen1992}. As a consequence, the detection of any signal related to radial stellar vibrations looks unlikely if quark matter is unpaired.

Still, in the CFL phase the contributions to the bulk viscosity from the above processes are exponentially suppressed \cite{Alford-Schmitt2007}. The thermodynamic and hydrodynamic properties are rather determined by the massless superfluid phonons $\varphi$ and  thermally  excited  light  pseudo-Nambu-Goldstone bosons. The contribution to bulk viscosity  from  phonons  alone ($\varphi \leftrightarrow \varphi + \varphi$) has  been  calculated  in \cite{manuel2007} (see also \cite{Escobedo} for a more recent analysis). For $T \gtrsim 0.1$ MeV and a typical oscillation period $\tau = 1$ ms, the  resulting  transport  coefficient  $\zeta^{\varphi}$  is several orders of magnitude smaller than for unpaired quark matter.

However, depending on the poorly known value for $\delta m \equiv  m_{K^0} - \mu_{K^0}$, the  dominant  contribution to the bulk viscosity may come from weak equilibrium processes involving the neutral kaon $K^0$ and the bosons $\varphi$, e.g. $K^0 \leftrightarrow \varphi + \varphi$ and $K^0  + \varphi \leftrightarrow \varphi$ \cite{Alford-Braby2007,Alford-Schmitt2007}. For oscillations with a timescale of milliseconds and temperatures above a few MeV, the bulk viscosity $\zeta^{K^0}$  can become larger than for unpaired quark matter (see \cite{Alford-Schmitt2007} and references therein). However,  $\zeta^{K^0}$ is much less than that of unpaired quark matter for low temperatures. For example, for $T \sim 0.1$ MeV and $\delta m = 0.1$ MeV,  $\zeta^{K^0}$ is more than six orders of magnitude smaller than $\zeta$ for unpaired matter. Since the dissipation time scale is $\tau \propto \zeta^{-1}$ \cite{Sawyer1989,madsen1992} we expect that radial oscillations of CFL stars would last for several seconds and are potentially observable after a catastrophic event involving the quark star.

Additionally, we must keep in mind that the spherical symmetry is broken if the star is subject to rotation causing a non-zero varying quadrupole moment that generates gravitational radiation \cite{Chau1967}. This energy sink has a characteristic timescale $\sim$ $10^3$ P$^4$ yr, where P is the period of rotation in seconds. Thus, while the persistence of primordial radial pulsations should not be expected, some fingerprints of radial oscillation modes could emerge in future observations  of violent transient phenomena.

\section{Acknowledgements}
C. V\'asquez Flores acknowledges the financial support received from UFABC and FAPESP. G. Lugones
acknowledges the financial support received from FAPESP and CNPq.

\end{document}